\documentclass[%
 reprint,
superscriptaddress,
 amsmath,amssymb,
 aps,
]{revtex4-1}

\usepackage{graphicx}
\usepackage{dcolumn}
\usepackage{bm}
\usepackage[version=3]{mhchem}
\usepackage{url}
\usepackage{color,soul}

\begin{document}

\preprint{APS/123-QED}

\title{Confinement effect on solar thermal heating process of TiN solutions}

\author{Anh D. Phan}
\affiliation{Faculty of Materials Science and Engineering, Phenikaa Institute for Advanced Study, Phenikaa University, Hanoi 100000, Vietnam}
\email{anh.phanduc@phenikaa-uni.edu.vn}
\affiliation{Faculty of Information Technology, Artificial Intelligence Laboratory,  Phenikaa University, Hanoi 100000, Vietnam}%
\affiliation{Department of Nanotechnology for Sustainable Energy, School of Science and Technology, Kwansei Gakuin University, Sanda, Hyogo 669-1337, Japan}

\author{Nam B. Le}
\affiliation{School of Engineering Physics, Hanoi University of Science and Technology, 1 Dai Co Viet, Hanoi, Vietnam}
\email{nble@mail.usf.edu}
\author{Nghiem T. H. Lien}
\affiliation{Institute of Physics, Vietnam Academy of Science and Technology, 10 Dao Tan, Ba Dinh, Hanoi 10000, Vietnam}
\author{Lilia M. Woods}
\affiliation{Department of Physics, University of South Florida, Tampa, Florida 33620, United States}
\author{Satoshi Ishi}
\affiliation{International Center for Materials Nanoarchitectonics, National Institute for Materials Science, Tsukuba 305-0044, Japan}
\author{Katsunori Wakabayashi}
\affiliation{Department of Nanotechnology for Sustainable Energy, School of Science and Technology, Kwansei Gakuin University, Sanda, Hyogo 669-1337, Japan}

\date{\today}

\begin{abstract}
We propose a theoretical approach to describe quantitatively the heating process in aqueous solutions of dispersed TiN nanoparticles under solar illumination. The temperature gradients of solution with different concentrations of TiN nanoparticles are calculated when confinement effects of the container on the solar absorption are taken into account. We find that the average penetration of solar radiation into the solution is significantly reduced with increasing the nanoparticle concentration. At high concentrations, our numerical results show that photons are localized near the surface of the solution. Moreover, the heat energy balance equation at the vapor-liquid interface is used to describe the solar steam generation. The theoretical time dependence of temperature rise and vaporization weight losses is consistent with experiments. Our calculations give strong evidence that the substantially localized heating near the vapor-liquid interface is the main reason for the more efficient steam generation process by floating plasmonic membranes when compared to randomly dispersed nanoparticles. The validated theoretical model suggests that our approach can be applied towards new predictions and other experimental data descriptions.
\end{abstract}

\pacs{Valid PACS appear here}
\maketitle


\section{Introduction}
Nanoplasmonics is a rapidly growing field in the areas of photothermal therapy \cite{2,3}, plasmonic-based solar cells \cite{5}, and high-efficiency photovoltaics \cite{6}. Plasmonic materials are typically associated with metals or metal-like materials which have negative real part of their permittivity in a given range of wavelength and models of their dielectric functions include a Drude term. Currently, most plasmonic systems are designed using noble metals such as gold or silver. The plasmonic properties are strongly dependent on size, shape, and structure \cite{2,3,5,6,45}, thus they are easily tunable. However, the low-melting-point property of noble metal nanoparticles restricts thermal-related applications, which operate at high temperatures. In this context, transition metal nitrides, such as TiN, have been proposed as alternative plasmonic materials for replacing conventional materials in nanoplasmonics \cite{22,7,24}. 

The advantages of TiN are stable thermodynamic properties at high temperatures, easier fabrication, and high affinity to existing standard silicon devices, and lower cost of production, which is important for devices at larger scales \cite{25,26}. Although the ceramic material has similar optical spectrum in the near infrared and visible regimes to the gold counterpart \cite{46,47,48}, its real permittivity has a much smaller magnitude due to smaller carrier concentration than gold or silver. In addition, the optical properties of TiN can be tuned easier by changing the processing conditions \cite{44}. 

The enhanced optical absorption capabilities of plasmonic nanoparticles can also be exploited for solar energy harvesting applications \cite{29,30}. Besides the storage of solar energy or its conversion into electricity, it is possible to directly transfer the solar energy into solutions to heat up and vaporize the solution. The solar steam generation process can decrease salinity levels of seawater and produce clean water \cite{28}. 

In solar steam generation and solar energy conversion, theoretical understanding for quantitative prediction of solar-induced temperature increase and vaporized weight loss has been lacking. A simple theoretical analysis using the heat transfer equation was proposed by Neumann et al. \cite{41} to understand the bubble formation surrounding a single gold nanoshell in the heating process in a solution containing the nanoparticles. However, the analysis provides only qualitative explanations for the phenomenon of temperature increase and bubble generation. Other studies are the heat transfer models associated with finite element analysis to calculate: (i) temperature gradient of a plasmonic nanostructure deposited on a glass substrate \cite{42} and (ii) the time evolution of temperature in a nanoparticle solution \cite{43} under simulated solar irradiation. 

Recently, we introduced a new plasmonic heating model for the time-dependent temperature increase of the gold-nanoshell solution under solar irradiation \cite{15}. This model assumed homogeneous temperature in the solution at given time during the heating process. Although the numerical results in Ref. \cite{15} agree well with experimental data at low concentrations of the nanoshells, the model overestimates the absorbed solar energy at high concentrations.

In this work, we theoretically investigate the application of TiN nanoparticles to solar energy harvesting. A theoretical approach is proposed to calculate the time-dependent thermal gradient and describe evaporation processes for a wide range of concentrations. It is demonstrated that the numerical results can agree with experimental data without introducing any adjustable parameters, which broadens the applicability of this model in practice. The theory-experiment agreement is also used to evaluate our physical description for the confinement effects of the solution container on the photothermal conversion.

\section{Theoretical background}
We consider the solar thermal heating and steam generation of an aqueous solution of TiN nanoparticles. The volume of solution is $V_0$, and the radius of nanoparticle is $R$. As electromagnetic radiation passes through these systems, the nanoparticles absorb the energy of radiation and dissipate into thermal energy to heat up the liquid medium. The sunlight absorption and photothermal conversion is described using Mie scattering theory and heat transfer equations. Here, we ignore radiative and convective heat losses in the solution. Since the typical plasmonic photothermal experiments are conducted in very dilute solutions, even the densest TiN nanoparticle concentration in Ref. \cite{22} ($10^{-1}$ vol $\%$) has an average nanoparticle separation of $\sim 32R$. Thus, all nanoparticles can be considered to be randomly dispersed and completely isolated in the solution. 

To calculate the spatial temperature profile in the solution of TiN nanoparticles under solar illumination, we adopt an analytic solution of a heat energy balance equation of thermal response of a semi-infinite substrate. In the case of no heat loss and instantaneous absorption optical energy during a heating process, the change of the temperature distribution function at coordinate ($x$, $y$, $z$) of a metal slab obeys the governing equation \cite{31}
\begin{eqnarray}
\frac{\partial^2 \Delta T}{\partial x^2} + \frac{\partial^2 \Delta T}{\partial y^2} &+& \frac{\partial^2 \Delta T}{\partial z^2}  -\frac{1}{\kappa}\frac{\partial \Delta T}{\partial t} \nonumber\\ &=& \frac{I(1-\mathcal{R})\alpha}{K}e^{-\beta^2(x^2+y^2)}e^{-\alpha z},
\label{eq:1}
\end{eqnarray}
where $z$ is the depth direction, $I$ is the laser intensity at the wavelength $\lambda$, and $\kappa = K/\rho c$ is the thermal diffusivity with $K $ being the thermal conductivity of water. Also, $\mathcal{R}$ is the reflectivity of the medium, $\beta$ is the inverse of the laser spot radius, and $\alpha$ is the absorption coefficient of the metal slab. Taking the Laplace transform of the above equation in $t$ gives
\begin{eqnarray}
\bigtriangledown^2\bar{\Delta T}(x,y,z,p) &-&\frac{p}{\kappa}\frac{\partial \bar{\Delta T}(x,y,z,p)}{\partial t} \nonumber\\ &=& \frac{I(1-\mathcal{R})\alpha}{pK}e^{-\beta^2(x^2+y^2)}e^{-\alpha z}. 
\label{eq:1-1}
\end{eqnarray}

Since the analytical solution of Eq. (\ref{eq:1-1}) is even with respect to $x$ and $y$, one can obtain the Laplace transformed solution by using the standard Fourier integral theory \cite{31}
\begin{eqnarray}
\bar{\Delta T}(x,y,z,p) &=& \frac{I(1-\mathcal{R})\alpha \kappa}{\pi^2 \beta^2 pK}\int_0^{\infty}\int_0^{\infty}\int_0^{\infty}\frac{e^{-(l^2+s^2)/4\beta^2}}{\kappa(l^2+s^2+v^2)+p} \nonumber\\
&\times& \left( \frac{\alpha}{\alpha^2+v^2} \right)\cos(lx)\cos(sy)\cos(vz)dldsdn.\nonumber\\
\label{eq:1-2}
\end{eqnarray}

Then the spatial and temporal temperature changes are obtained by simply taking the inverse Laplace transform of Eq. (\ref{eq:1-2}). The analytical expression of $\Delta T(x,y,z,t)$ is
\begin{widetext}
\begin{eqnarray}
\Delta T(x,y,z,t)=\frac{I(1-\mathcal{R})\alpha}{2\rho c}\int_0^t \exp\left(-\frac{\beta^2(x^2+y^2)}{1+4\beta^2\kappa t'} \right)\frac{e^{\alpha^2\kappa t'}}{1+4\beta^2\kappa t'}\left[e^{-\alpha z}\ce{erfc}\left(\frac{2\alpha\kappa t'-z}{2\sqrt{\kappa t'}} \right)+ e^{\alpha z}\ce{erfc}\left(\frac{2\alpha\kappa t'+z}{2\sqrt{\kappa t'}} \right)\right]dt',
\label{eq:14}
\end{eqnarray}
\end{widetext}

By considering an effective absorption coefficient incorporating the  effects of the water and dispersed nanoparticles, Eq. (\ref{eq:14}) can be extended to the dilute solution. The experiments of Ref. \cite{22} are conducted in a beaker (diameter $D=4$ cm and height $H=1.6$ cm). Thus, the effective absorption coefficient can be modeled as
\begin{eqnarray}
\alpha(\omega) =\left\{ \begin{array}{rcl} 
\alpha_w(\omega) + N Q_{ext} & \mbox{for} & 1/\alpha(\omega) \leq H\\
1/H & \mbox{for} & 1/\alpha(\omega) > H
\end{array}\right.
\label{eq:16}
\end{eqnarray}
where $\alpha_w$ is the absorption coefficient of water \cite{23}, $N$ is the number of particles per unit volume in the solution, and $Q_{ext}$ is the extinction cross section calculated using Mie scattering theory \cite{1}. 

The first relation in Eq. (\ref{eq:16}) reflects Beer-Lambert law for the nanoparticle solution in a bulk state  with no confinement effects. The second relation corresponds to a confinement condition of trapped photons inside a beaker that is thermally isolated around its circumference as reported in Ref. \cite{22}. By assuming that the photon mean free path is limited by the height of the beaker, the absorption coefficient can be taken as the inverse of $H$ as the second relation of Eq. (\ref{eq:16}).

In calculation of Mie scattering, the host medium is water with the dielectric function $\varepsilon_h(\omega) \approx 1.77$. The TiN nanoparticles are nonmagnetic. The complex dielectric function of TiN is described using a generalized Drude-Lorentz model, which is fitted by experimental data for thin film materials \cite{11}, 

\begin{eqnarray}
\varepsilon(\omega)=\varepsilon_{\infty}-\frac{\omega_{p}^2}{\omega(\omega+i\Gamma_D)}+\sum_{j=1}^2\frac{\omega_{L,j}^2}{\omega_{0,j}^2-\omega^2-i\gamma_j\omega},
\label{eq:3}
\end{eqnarray}
where $\varepsilon_{\infty}=5.18$ is the permittivity at infinite frequency. Also, $\omega_{p} \approx 7.38$ eV, and $\Gamma_D \approx 0.26$ eV \cite{11} are the plasma frequency and the Drude damping parameter at the $j$th mode, respectively.  The Lorentz oscillator strengths are $\omega_{L,1}\approx 6.5$ \ce{eV} and $\omega_{L,2}=1.5033$ \ce{eV}, the Lorentz energies are $\omega_{0,1}=4.07$ \ce{eV} and $\omega_{0,2}=2.02$ \ce{eV}, and the Lorentz damping parameters are $\gamma_1 = $ 1.42 \ce{eV} and $\gamma_1 = $ 1.42 \ce{eV} \cite{11}. 

The calculated optical spectrum of TiN nanoparticle is relatively close to that of the gold counterpart \cite{46,47,48}. For nanoparticles of 50-nm radius dispersed in water, our numerical calculations show that the resonance wavelength of the surface plasmon of TiN sphere is at about 600 nm, while that of gold sphere is located at approximately 530 nm. The results are in accordance with prior work \cite{46}. Consequently, one expects that the response of solution of TiN nanoparticles under illumination of electromagnetic fields to behave qualitatively in the same manner as that of gold nanoparticles. Moreover, it also suggests the  Eq. (\ref{eq:14}) is still valid when applied to TiN solutions.

Since the considered solution of TiN nanoparticles is exposed under simulated sunlight and the reflectivity of solution environment is relatively small, we can approximate $\mathcal{R} \approx 0$ and rewrite Eq. (\ref{eq:14}) to be

\begin{widetext}
\begin{eqnarray}
\Delta T(x,y,z,t)=\int_0^t dt'\int_{300}^{1500}d\lambda \frac{E_\lambda\alpha}{2\rho c}\exp\left(-\frac{\beta^2(x^2+y^2)}{1+4\beta^2\kappa t'} \right)\frac{e^{\alpha^2\kappa t'}}{1+4\beta^2\kappa t'}\left[e^{-\alpha z}\ce{erfc}\left(\frac{2\alpha\kappa t'-z}{2\sqrt{\kappa t'}} \right)+ e^{\alpha z}\ce{erfc}\left(\frac{2\alpha\kappa t'+z}{2\sqrt{\kappa t'}} \right)\right], \nonumber\\
\label{eq:17}
\end{eqnarray}
\end{widetext}
where $E_\lambda$ is the solar spectral irradiance of AM1.5 global solar spectrum. Since the electromagnetic radiation of the XES-40S1 solar simulator is in the 300-1500 nm wavelength range, the lower and upper limit of the integral in Eq. (\ref{eq:17}) are 300 and 1500 nm. We note that the wavelength range of operation depends on the particular devise used as a sunlight simulator, and the upper limit can reach 4000 nm in particular devices. We assume that the spot size is infinite, thus $\beta = 0$. In the solution, $\rho = 1000$ $kg/m^3$ is approximately the mass density of water and $c=4200$ $J/kg/K$ is the specific heat of water.

The derivation of Eq. (\ref{eq:14}) is based on an important assumption that the substance absorbs instantaneously the laser light energy. The assumption is very plausible in the case of laser light since the intensity is high. The intensity of solar illumination, however, is approximately 100 $\ce{mW/cm^2}$ which is much smaller than the laser intensity, meaning that this assumption may need to be re-evaluated. Nevertheless, even in this case Ref. \cite{41} has shown that the instantaneous light absorption leads to reasonable results for the description of solar illumination of nanoparticle solutions.

After obtaining the thermal gradient by using Eq. (\ref{eq:17}), the spatial distribution of temperature rise is averaged for different concentrations to compare with experimental data. The averaged increase of temperature is
\begin{eqnarray}
\Delta T_{ave}(t)=\frac{1}{H}\int_0^{H}\Delta T(x,y,z,t)dz.
\end{eqnarray}

At the liquid-vapor interface, we assume that the convective heat transfer is approximately cooled down by evaporation. If there is no thermal dissipation into the environment, the weight loss in the vaporization process is essentially determined by the temperature discrepancy between the solution surface and surrounding medium. Thus according to the energy balance equation, one has
\begin{eqnarray}
\frac{\pi D^2}{4}h_c\Delta T_{surface}(t) = -L\Delta m, 
\label{eq:8}
\end{eqnarray}
where $\Delta T_{surface}(t)$ is the time dependence of temperature difference at the liquid-steam interface, $h_c$ is the convection heat transfer coefficient, and $\Delta m$ is the vaporized weight of water. $h_c$ can vary from 50 to 10000 \ce{W/(m^2K)} depending on the materials \cite{12}. For example, for free (natural) convection, $h_c = 500$ \ce{W/(m^2K)}, while for steam media $h_c=10000$ \ce{W/(m^2K)}. In this work, we consider the heat transport of hot water in a breaker at the liquid-steam interface. Since the vaporization process occurs under the simulated sunlight illumination, thus the coefficient is approximately 10000 \ce{W/(m^2K)}. Also,  $L=2.26\times 10^6$ J/kg is the specific heat of evaporation of water. 

\section{Numerical results and discussions}
\begin{figure}[htp]
\center
\includegraphics[width=9cm]{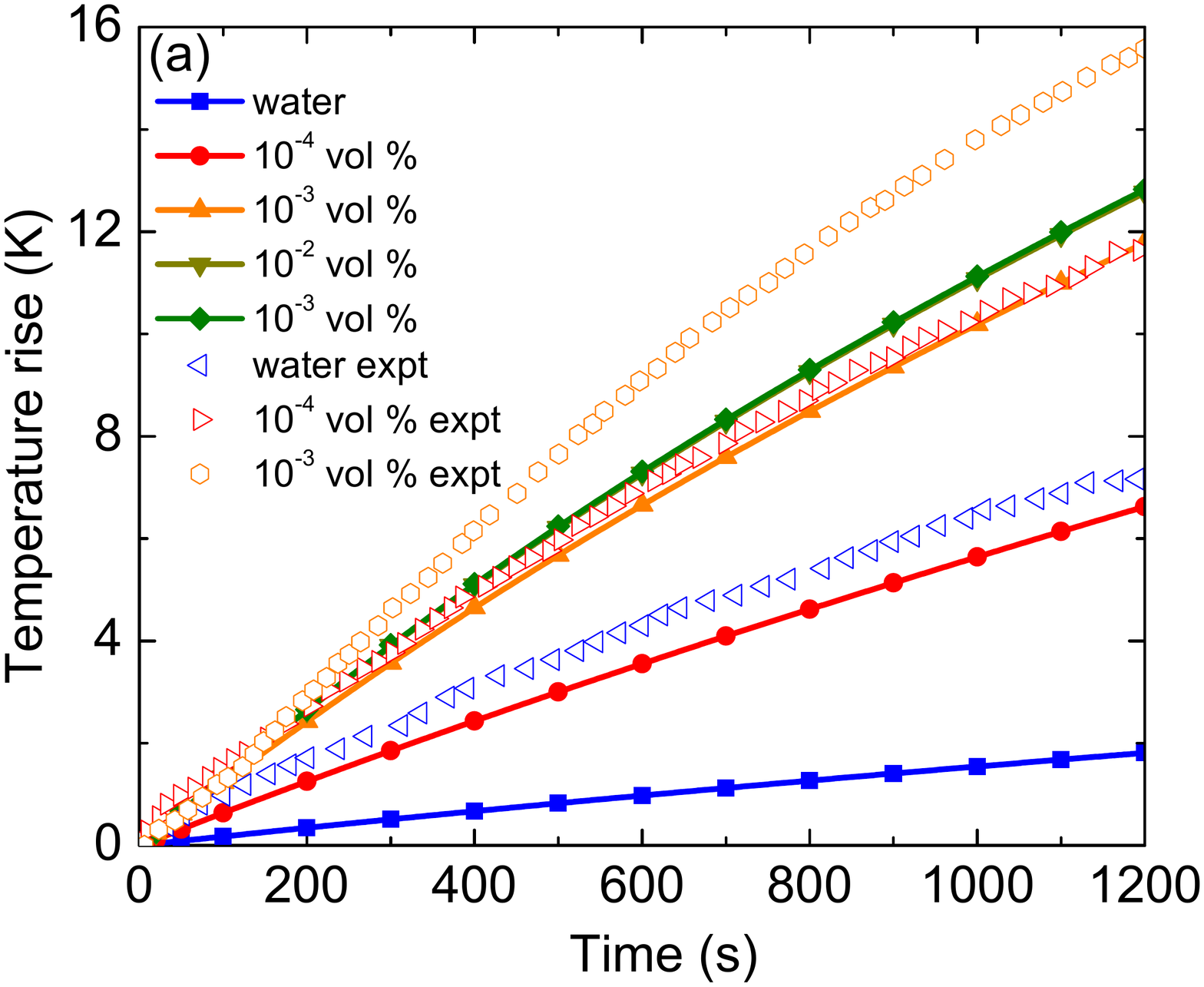}
\includegraphics[width=9cm]{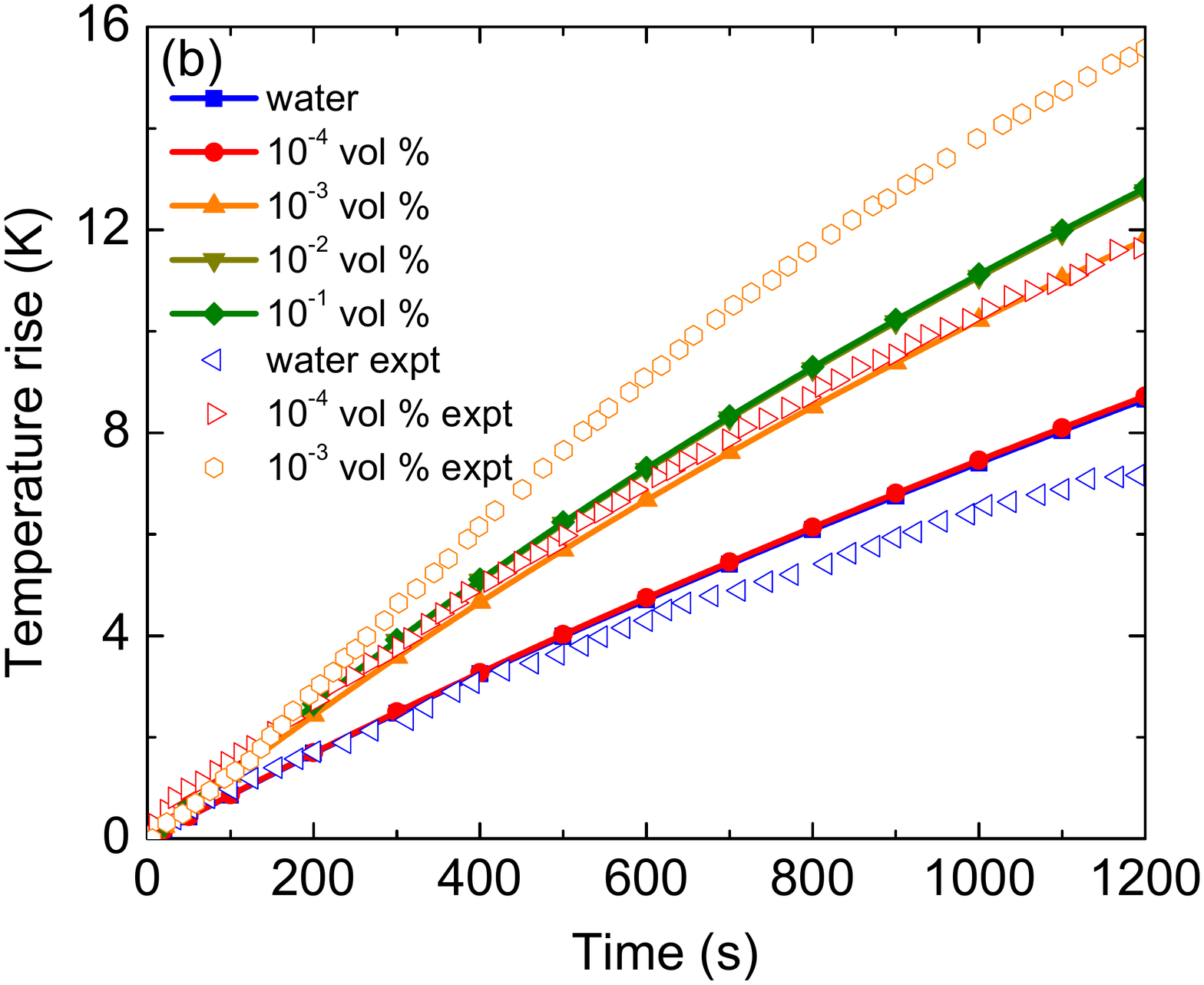}
\caption{\label{fig:4}(Color online) Time dependence of the average temperature rise for nanoparticle solutions with different concentrations of TiN nanoparticles calculated using Eq. (\ref{eq:17}) with (a) no confinement effect, and (b) confinement effect (Eq.(\ref{eq:16})). Curves correspond to theoretical calculations and scatter points are experimental data taken from Ref. \cite{22}.}
\end{figure}

Figure \ref{fig:4} shows the comparison between our theoretical calculations and  experimental data for the time-dependent temperature increase of water and solutions of TiN nanoparticles. The experimental data is taken from Ref. \cite{22}. The numerical results without the confinement effects are shown in Fig. \ref{fig:4}a. One finds that in this case $\Delta T_{ave}(t)$ is underestimated, if we compare with the experimental data for pure water, and TiN nanofluid of $10^{-4}$ and $10^{-3}$ vol $\%$. 

A better agreement with the experimental data is obtained by taking into account confinement effects, as shown from Fig. \ref{fig:4}b. However, the theoretical curve corresponding to TiN nanofluid of $10^{-4}$ vol $\%$ almost overlaps with that of water. The result suggests that the contribution of this amount of nanoparticles to the effective absorption is minor in our model. The TiN nanofluid with concentration of $10^{-4}$ vol $\%$ is an extremely dilute solution, thus the simulated sunlight passes through the entire system rather easily. 

At higher concentrations ($\geq 10^{-3}$ vol $\%$), the calculations in Fig. \ref{fig:4}b are in a good agreement with the experiments without any adjustable parameters. These findings reveal that the average penetration depth of photons, $1/\alpha(\omega)$, is reduced and less than the solution height. This also indicates that the electromagnetic field of the incident light on average is absorbed before reaching the bottom of the breaker. Additionally, our proposed assumption for the confinement effect is reasonable and gives a good qualitative physical picture for the optical absorption. The time dependence of temperature at concentration of $10^{-2}$ vol $\%$ is slightly above that of $10^{-3}$ vol $\%$ but is nearly unchanged with increase of the concentration up to $10^{-1}$ vol $\%$. One can see similar trends in the experiment \cite{22}.  Thus, this model provides good predictions for new experiments.

\begin{figure*}[htp]
\center
\includegraphics[width=17.5cm]{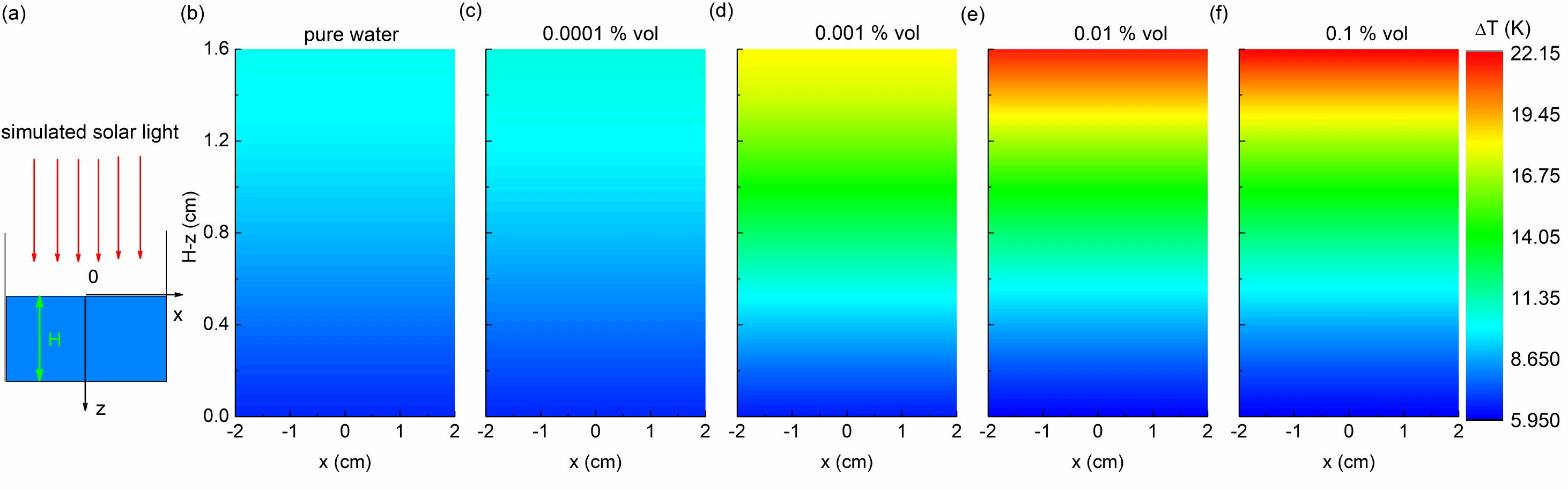}
\caption{\label{fig:1}(Color online) (a) Schematic illustration of the solar-irradiated beaker. Spatial contour plots of the temperature rise in Kelvin's unit after 1200s under solar illumination calculated using Eq.(\ref{eq:17}) and taking $y=0$ in the solution of TiN nanoparticles having a concentration of (b) 0 vol $\%$, (c) 0.0001 vol $\%$, (d) 0.001 vol $\%$, (e) 0.01 vol $\%$, and (f) 0.1 vol $\%$, respectively. }
\end{figure*}

Figure \ref{fig:1}a schematically shows the system to calculate its temperature profile using Eqs. (\ref{eq:16}) and (\ref{eq:17}). Figures \ref{fig:1}b-f show the density plots of $\Delta T$ for different concentrations of nanoparticle illuminated by solar light after $t=1200$ s. In the case of pure water, the local temperature rise at 0.8 cm below the air-liquid interface (the center of the solution) is approximately 8.87 $K$ which is relatively less than $\Delta T_{ave} \approx 9.37$ $K$. The numerical results in Fig. \ref{fig:1}c show that for a very dilute solution with $10^{-4}$ vol $\%$, the TiN nanoparticles have a minor effect on the temperature distribution. At the hottest spot area of the system ($z = 0$), the calculated temperature increase at the surface is 10.45 $K$, which is different than the experimentally reported value of $\Delta T(t=1200s) = 11.63$ $K$ for a similar system \cite{22}. At a higher concentration of nanoparticles ($10^{-3}$ vol $\%$), the local temperature rise (up to 14.5 $K$) near the air-liquid interface agrees very well with the experiments in Ref. \cite{22}. Our calculations indicate that an increase of nanoparticle concentration greater than 0.01 vol $\%$ does not significantly change the temperature profile in the solution, which is also consistent with the measurements in Ref. \cite{22}. The thermal gradient is very broad and the temperature deviation between the top and bottom of the glass beaker is significant. The light-to-heat conversion process is highly localized near the surface and rapidly decays towards the bottom of the container. This is attributed to a substantial reduction of the optical mean free path with the increase of TiN nanoparticle concentration in the solution.

\begin{figure}[htp]
\center
\includegraphics[width=9cm]{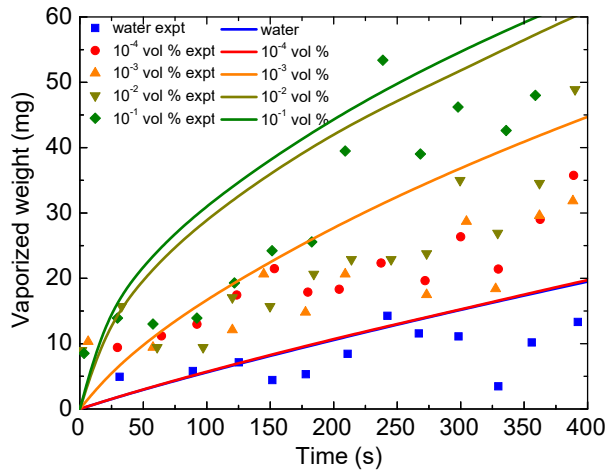}
\caption{\label{fig:3}(Color online) Theoretical (solid lines) and experimental (data points) vaporized weight of water and TiN nanofluid with several concentrations as a function of time.}
\end{figure}

It is further instructive to study the time-dependent weight change of the water/TiN nanoparticles system in the heat conversion process, which gives important insight in the solar vapor generation. Figure \ref{fig:3} shows the time dependence of weight changes for solutions with $10^{-4}$, $10^{-3}$, $10^{-2}$, and $10^{-1}$ vol $\%$ concentration calculated by using Eq. (\ref{eq:8}). The theoretical curves, particularly for lower concentrations ($\leq$ $10^{-4}$ vol $\%$), agree quantitatively with experimental data in Ref. \cite{22}. It is hard to see a mathematical form of the weight loss as a function of time based on experimental data points since they relatively fluctuates. However, other researchers have reported less noisy measurements for solar steam generation \cite{41,17,18,19,50}. One can see a rather linear time dependence of weight water evaporation in Ref. \cite{41,17,18,19,50}, which also agrees with our theoretical analysis. At higher concentrations ($\geq$ 0.001 vol $\%$), Eq. (\ref{eq:8}) suggests the time dependencies of the vaporized weight are unchanged with increase of the TiN particle density.

Equation (\ref{eq:8}) contains dependence of mass changes to the temperature deviation between surface and vapor medium. This is helpful to consider different systems for solar steam generation, which have floating particle membranes at the air-liquid interface and submerged particle membranes on the bottom of the glass beaker \cite{50}. Our model suggests that floating photothermal agents on the solution surface provides larger temperature difference at the air-liquid interface than distributing randomly them in the solution, because the densification significantly increases the collective heating effects. Furthermore, floating photothermal materials on the water surface can reduce heat dissipation compared to randomly disperse the photothermal agents in the water.

\section{Conclusions}
A theoretical approach has been proposed to describe the plasmonic heating process and solar steam generation in a solution of dispersed TiN nanoparticles under solar illumination. We have calculated the temperature gradient in a finite solution from the analytical solution of the heat transfer differential equation. Increasing the number of TiN nanoparticles leads to an increase of absorbed energy by the nanoparticles, resulting in a steeper increase of the time-dependent temperature profile. The penetration depth of the simulated sunlight in water is reduced at higher concentrations and much less than the height of solutions. We have examined the validity of the approach by calculating temperature profiles compared with reliable experimental data, which is instrumental in evaluating the advantages and limitations of this model. These numerical results also suggest that the local heating at the solution surface is significantly enhanced when plasmonic materials are localized near the liquid-vapor interface. The densification of nanoparticles increases the temperature difference between air and liquid at the interface compared to solutions of random nanoparticle distribution. From the predicted temperature increase profiles, we have theoretically estimated the vaporized weight by assuming the evaporation process is strongly dependent on the temperature deviation between the solution and environment. The approach provides a guidance to design efficient solar steam generation applications.

\section*{Conflicts of interest}
There are no conflicts to declare.

\section*{Acknowledgements}
This research is funded by Vietnam National Foundation for Science and Technology Development (NAFOSTED) under grant number 103.01-2017.63. L. W. acknowledges support from the US Department of Energy under Grant No. DE-FG02-06ER46297. This work was supported by JSPS KAKENHI Grant Numbers JP19F18322 and JP18H01154.

\end{document}